\documentclass[aps,preprint,showpacs,showkeys]{revtex4}
\begin{document}
\preprint{\tt quant-ph/0408127 }
\title{Lorentz-invariant Bell's inequality}
\author{Won Tae Kim}
\email{wtkim@sogang.ac.kr}
\author{Edwin J. Son}
\email{eddy@sogang.ac.kr}
\affiliation{Department of Physics and Basic Science Research Institute,\\
         Sogang University, C.P.O. Box 1142, Seoul 100-611, Korea}
\date{\today}

\begin{abstract}
We study Bell's inequality in relation to the Einstein-Podolsky-Rosen
paradox in the relativistic regime. For this purpose, a
relativistically covariant analysis is used in the calculation
of the Bell's inequality, which results in the maximally violated
Bell's inequality in any reference frame.  
\end{abstract}
\maketitle

Since Einstein, Podolsky, and Rosen (EPR)~\cite{epr} have proposed 
an intriguing gedanken experiment, there have been great
efforts to shed light on the nonlocality of quantum 
mechanics. Violation of the Bell's inequality~\cite{bell}, especially
as seen from the Bohm spin version, may indicate 
its nonlocality characteristically even though it seems to be
contradictory to the special theory of relativity based on the locality. 
Recently, this inequality and entanglement in nonrelativistic 
quantum mechanics have been promoted to the moving observer in the
relativistic limit by a number of authors~\cite{czachor,am,tu,almh,le,pst};
however, they have not always reached the same conclusion for
Bell's inequality. Some of them~\cite{am,tu} have claimed  that violation
of Bell's inequality can be recovered by the unitary Wigner rotation under
a Lorentz boost even though the amount of violation decreases depending
on the boost velocity. On the other hand, others~\cite{almh} asserted 
that Bell's inequality is
satisfied in the ultrarelativistic limit by using the relativistic
spin operator suggested by Czachor~\cite{czachor}.

In order to study the transformation of maximally entangled states under
the action of Lorentz transformations, the authors in
Refs.~\cite{am,tu} considered only the changes of the states. In
Ref.~\cite{czachor}, however, the nonrelativistic
singlet state average is relativistically generalized by defining spin
via the relativistic center-of-mass operator. Using this relativistic
spin observable, its expectation values for the Bell states under
Lorentz boost have been evaluated in Refs.~\cite{almh,le},
and it has been shown that special
relativity imposes severe restrictions on the transfer of information
between distant systems and the implications of the fact that quantum
entropy is not a Lorentz-covariant concept by using
positive-operator-valued measures in Ref.~\cite{pst}.

In this Brief Report, we would like to study Bell's inequality in the
moving frame by using the relativistically invariant Bell observable.
In principle, consistent results should come from the
relativistically covariant analysis. By using a consistent
transformation, in both the Bell states and the observable, we
naturally obtain an observer-independent Bell's inequality, so that it
is maximally violated as long as it is violated maximally in the rest frame.
Finally, adiscussion will be given 
of the frame-independent result compared with previous works.    

For a single spinning relativistic massive particle, 
all unitary irreducible representations of the Poincar\'e
group~\cite{ohnuki} rely on Wigner's idea that the quantum states can
be formulated directly without the use of the wave equation
\cite{wigner}. Following Wigner's approach,  
the Lorentz transformation $\Lambda$ induces a unitary transformation
\footnote{In general, a particle has no definite momentum. The
  validity of the transformation rules arises from the assumption that
  particle states have sharp momenta, namely are momentum
  eigenstates. If this condition is not met, the Wigner matrix
  generates entanglement between momentum and spin degrees of freedom,
  thus resulting in a change of entropy of the spin state of the
  system~\cite{pst}.} on
a particle state $|p,\sigma\rangle$ with a four-momentum $p=(p^0,\vec{p})$ and
a spin $\sigma$ as~\cite{ryder}
\begin{equation}
U(\Lambda) |p,\sigma\rangle = \sum_{\sigma'} {\mathcal D}_{\sigma'\sigma} (W(\Lambda,p)) |\Lambda
p,\sigma'\rangle, \label{rep:state}
\end{equation}
where $L(p)$ is the Lorentz transformation, which makes a rest particle
move with the momentum $p$, i.e., $p^\mu =
L^\mu_{\phantom{\mu}\nu} (p) k^\nu$, 
$k^\mu = (m, 0, 0, 0)$ for $\mu,
\nu = 0,1,2,3$, and ${\mathcal D} (W)$ is the representation of Wigner's
little group element given by
$ W(\Lambda,p) = L^{-1}(\Lambda p) \Lambda L(p)$.
Note that the speed of light $c$ and the Planck constant $\hbar$ are
set to be 1 throughout this paper, but they will be written
explicitly when necessary.

We will consider two reference frames in this work: one is the rest
frame $S$ and the other is the moving frame $S'$ in which a particle
whose four-momentum $p$ in $S$ is seen as boosted with the velocity
$\vec{\beta}$. By setting the boost and particle moving directions
in the rest frame to be  $\hat{\beta} = \hat{x}$ and 
 $\hat{p} = \hat{z}$, respectively, 
the Wigner representation is found as
\begin{eqnarray}
{\mathcal D}(W(\Lambda,p)) = \left( \begin{array}{cc} \cos
    \frac{\Omega}{2} & -\sin \frac{\Omega}{2} \\ \sin \frac{\Omega}{2}
    & \cos \frac{\Omega}{2} \end{array} \right), \label{rep:wig}
\end{eqnarray}
where the Wigner angle $\Omega$ satisfies the relation
$\tan \Omega = \sinh \alpha \sinh \delta /(\cosh \alpha + \cosh
  \delta)$, $\cosh \alpha = \gamma = 1/\sqrt{1-\beta^2}$, and
  $\cosh \delta = p^0/m$.
In the relativistic limit, $\beta \to 1$ or $\alpha \to \infty$,
the Wigner angle $\Omega$ becomes $\pi/2$ when the energy of the
particle is very high, $p^0/m \to \infty$ ($\delta \to \infty$),
while it approaches zero for low energy, $p^0/m \to 1$ ($\delta
\to 0$).

We now consider two spin-1/2 particles moving in the opposite
directions $\hat{p}$ and $-\hat{p}$ with the same energy and speed 
in $S$. Then, the entangled Bell states~\cite{nc} in $S$ are given by
\begin{eqnarray}
|\Psi_{++-- }^{(+)}\rangle &=& \frac{1}{\sqrt2} \left[ |p,\frac12\rangle \otimes
  |{\mathcal P}p,\frac12\rangle + |p,-\frac12\rangle \otimes
  |{\mathcal P}p,-\frac12\rangle \right] = |p,{\mathcal P}p\rangle \otimes
|\Phi_{++--}^{(+)}\rangle, \label{psi00} \\
|\Psi_{++--}^{(-)}\rangle &=& \frac{1}{\sqrt2} \left[ |p,\frac12\rangle \otimes
  |{\mathcal P}p,\frac12\rangle - |p,-\frac12\rangle \otimes
  |{\mathcal P}p,-\frac12\rangle \right] = |p,{\mathcal P}p\rangle \otimes
|\Phi_{++--}^{(-)}\rangle, \label{psi01} \\
|\Psi_{+--+}^{(+)}\rangle &=& \frac{1}{\sqrt2} \left[ |p,\frac12\rangle \otimes
  |{\mathcal P}p,-\frac12\rangle + |p,-\frac12\rangle \otimes
  |{\mathcal P}p,\frac12\rangle \right] = |p,{\mathcal P}p\rangle \otimes
|\Phi_{+--+}^{(+)}\rangle, \label{psi10} \\
|\Psi_{+--+}^{(-)}\rangle &=& \frac{1}{\sqrt2} \left[ |p,\frac12\rangle \otimes
  |{\mathcal P}p,-\frac12\rangle - |p,-\frac12\rangle \otimes
  |{\mathcal P}p,\frac12\rangle \right] = |p,{\mathcal P}p\rangle \otimes
|\Phi_{+--+}^{(-)}\rangle, \label{psi11}
\end{eqnarray}
where ${\mathcal P}$ is a parity operator satisfying ${\mathcal P}p =
(p^0,-\vec{p})$,
and $|\Phi_{++--}^{(+)}\rangle, |\Phi_{++--}^{(-)}\rangle, |\Phi_{+--+}^{(+)}\rangle$, and
$|\Phi_{+--+}^{(-)}\rangle$
are Bell bases in the rest frame. 
Note that $\pm$ in the Bell states represent spins up and
down, respectively. 
Then, the Lorentz boosted Bell states are found to be
\begin{eqnarray}
|\Psi_{++--}^{(+)'}\rangle &=& U(\Lambda) |\Psi_{++--}^{(+)}\rangle = |\Lambda p,\Lambda {\mathcal
  P}p\rangle \otimes \left[ \cos \Omega |\Phi_{++--}^{(+)}\rangle - \sin \Omega |\Phi_{+--+}^{(-)}\rangle
  \right], \label{rel:psi00} \\
|\Psi_{++--}^{(-)'}\rangle &=& U(\Lambda) |\Psi_{++--}^{(-)}\rangle = |\Lambda p,\Lambda {\mathcal
  P}p\rangle \otimes |\Phi_{++--}^{(-)}\rangle, \label{rel:psi01} \\
|\Psi_{+--+}^{(+)'}\rangle &=& U(\Lambda) |\Psi_{+--+}^{(+)}\rangle = |\Lambda p,\Lambda {\mathcal
  P}p\rangle \otimes |\Phi_{+--+}^{(+)}\rangle, \label{rel:psi10} \\
|\Psi_{+--+}^{(-)'}\rangle &=& U(\Lambda) |\Psi_{+--+}^{(-)}\rangle = |\Lambda p,\Lambda {\mathcal
  P}p\rangle \otimes \left[ \sin \Omega |\Phi_{++--}^{(+)}\rangle + \cos \Omega |\Phi_{+--+}^{(-)}\rangle
  \right] \label{rel:psi11}
\end{eqnarray}
by using the Wigner representation (\ref{rep:wig}). Note that
$|\Phi_{++--}^{(+)}\rangle$ and $|\Phi_{+--+}^{(-)}\rangle$ are rotated, while $|\Phi_{++--}^{(-)}\rangle$ and
$|\Phi_{+--+}^{(+)}\rangle$ are Lorentz invariant.  
So far, we have obtained the Lorentz transformed Bell states in terms
of the Wigner rotation, whose representation has been evaluated by
using the two-component spinor representation. 

In the rest frame, the spin
observable in the direction $\vec{a}$ is given by $\vec{a}
\cdot \vec{S}$, where $\vec{a}$ is a unit vector, $\vec{S} = (\hbar/2)
\vec{\sigma}$ is the spin operator, and the $\sigma^i$'s are
the Pauli matrices. Using the Pauli-Lubanski pseudovector $W_\mu = -
1/2 \epsilon_{\mu\nu\kappa\rho} J^{\nu\kappa} P^\rho$ with the
generators of the Poincar\'e group, $J^{\nu\kappa}$ and $P^\rho$
\cite{ryder}, the invariant expression measured by the
four-dimensional axis is assumed to be
\begin{eqnarray}
\hat{\mathcal O} (a) = \frac{2 a^\mu W_\mu}{mc\hbar},
\label{inv:ob}
\end{eqnarray}
where $a^\mu = (0,\vec{a})$ and $W^\mu =
(0,m\vec{S})$ in the rest frame, and $a^\mu a_\mu=1$. 
The spin vector and the axis should
be transformed by the appropriate transformation law. 
For the observable $\hat{\mathcal O}(a,b) = \hat{\mathcal O}(a)
\otimes \hat{\mathcal O}(b)$ acting on the Bell states in the rest frame, the
degree of violation of the Clauser-Horne-Shimony-Holt
inequality~\cite{chsh}, a variant of Bell's inequality, is measured by
\begin{equation}
C(a_1,a_2,b_1,b_2;\Psi) = \langle\hat{\mathcal O}(a_1,b_1)\rangle + \langle\hat{\mathcal
  O}(a_1,b_2)\rangle + \langle\hat{\mathcal O}(a_2,b_1)\rangle - \langle\hat{\mathcal
  O}(a_2,b_2)\rangle, 
\end{equation}
where the axes $a_1, a_2, b_1$, and $b_2$ are all four-vectors.
Then, the transformed expectation value
$\langle\Psi'|\hat{\mathcal O}'|\Psi'\rangle$ can be calculated by noting that
the observable transforms as $\hat{\mathcal O}'(a,b) = U(\Lambda) \hat{\mathcal O}(a,b)
U^{-1}(\Lambda) =  (2/mc\hbar)^2 a^\mu  b^\nu U(\Lambda) W_\mu \otimes
W_\nu U^{-1}(\Lambda)=
4 \vec{a} \cdot \vec{S}_{\mathrm R} \otimes \vec{b}
\cdot \vec{S}_{\mathrm R}$, where
$\vec{S}_{\mathrm R} = {\mathcal D}(W) \vec{S} {\mathcal
  D}^{-1}(W)$. 
By the use of Eq.~(\ref{rep:wig}), the transformation of the spin is 
rewritten by the transformation of the axes along with the following
relation:
\begin{eqnarray}
2 \vec{a} \cdot \vec{S}_{\mathrm R} &=& \vec{a} \cdot {\mathcal D}(W)
\vec{\sigma} {\mathcal D}^{-1}(W) \nonumber \\
&=& \left( \begin{array}{cc} a_z \cos \Omega -
    a_x \sin \Omega & a_z \sin \Omega + a_x \cos \Omega -
    i a_y \\ a_z \sin \Omega + a_x \cos \Omega + i a_y
    & - a_z \cos \Omega + a_x \sin \Omega \end{array} \right)
\nonumber \\
&=&2 \vec{a}_{\mathrm R} \cdot \vec{S},
\end{eqnarray}
which yields $\hat{\mathcal O}'(a,b) = \hat{\mathcal O}(
\vec{a}_{\mathrm R}, 
\vec{b}_{\mathrm R})$.
Then, under the Wigner rotation, the unit vectors $\vec{a}$ and
$\vec{b}$ are transformed as $\vec{a}_{\mathrm R} = (a_x \cos \Omega +
a_z \sin \Omega, a_y, -a_x \sin \Omega + a_z \cos \Omega)$ and 
$\vec{b}_{\mathrm R}   = (b_x \cos \Omega - b_z \sin \Omega, b_y, b_x \sin \Omega
+ b_z \cos \Omega)$. After some tedious calculations, 
the expectation values of the spin observable are calculated as
\begin{eqnarray}
\langle\Psi_{++--}^{(+)'}|\hat{\mathcal O}' (a,b)|\Psi_{++--}^{(+)'}\rangle &=&
a_x b_x - a_y b_y + a_z b_z, \label{exp:mod00} \\
\langle\Psi_{++--}^{(-)'}|\hat{\mathcal O}' (a,b)|\Psi_{++--}^{(-)'}\rangle &=&
-a_x b_x + a_y b_y + a_z b_z, \\
\langle\Psi_{+--+}^{(+)'}|\hat{\mathcal O}' (a,b)|\Psi_{+--+}^{(+)'}\rangle &=&
a_x b_x + a_y b_y - a_z b_z, \\
\langle\Psi_{+--+}^{(-)'}|\hat{\mathcal O}' (a,b)|\Psi_{+--+}^{(-)'}\rangle &=&
-a_x b_x - a_y b_y - a_z b_z. \label{exp:mod11}
\end{eqnarray}
Note that the expectation values are all invariant under the
Lorentz boost. Using the expectation values
(\ref{exp:mod00})--(\ref{exp:mod11}), it can be shown that the maximal
violation of Bell's inequality is maintained at any boost speed
$\beta$ and any particle speed $v_p$:
\begin{eqnarray}
C'(a_1,a_2,b_1,b_2;\Psi') = C(a_1,a_2,b_1,b_2;\Psi) = 2\sqrt2, \label{bell:mod}
\end{eqnarray}
where we chose the axes as
\begin{eqnarray}
& & \vec{a}_1 = (1/\sqrt2,-1/\sqrt2,0),\ \vec{a}_2 =
(-1/\sqrt2,-1/\sqrt2,0) \quad \textrm{ for } |\Psi_{++--}^{(+)'}\rangle, \\
& & \vec{a}_1 = (-1/\sqrt2,1/\sqrt2,0),\ \vec{a}_2 =
(1/\sqrt2,1/\sqrt2,0) \quad \textrm{ for } |\Psi_{++--}^{(-)'}\rangle, \\
& & \vec{a}_1 = (1/\sqrt2,1/\sqrt2,0),\ \vec{a}_2 =
(-1/\sqrt2,1/\sqrt2,0) \quad \textrm{ for } |\Psi_{+--+}^{(+)'}\rangle, \\
& & \vec{a}_1 = (-1/\sqrt2,-1/\sqrt2,0),\ \vec{a}_2 =
(1/\sqrt2,-1/\sqrt2,0) \quad \textrm{ for } |\Psi_{+--+}^{(-)'}\rangle,
\label{axes:singlet}
\end{eqnarray}
and $\vec{b}_1 = (0,1,0)$ and $\vec{b}_2 = (1,0,0)$ for
all Bell states. Of course, these sets of spin measurement axes are
equivalent to those giving maximal violation of Bell's inequality in the
nonrelativistic limit $\beta \to 0$.

The operator (\ref{inv:ob}) in the rest frame is in fact 
equivalent to the one defined in
Ref.~\cite{tu}. So, one might wonder what the difference is.
As shown in Ref.~\cite{tu}, it is possible to find the
direction for the maximal violation of Bell's inequality
(anticorrelation) in the relative motion after the Wigner rotation. 
It is true, however, as we found, that
the directions or axes should be transformed according to the Lorentz
transformation rule from the beginning.  
To rotate the axes arbitrarily in one's frame means that
one has to have individual apparatus for each EPR experiment. 
However, our physical system is unique so that the present physical
axes should be transformed according to the coordinate transformation rule. 
For example, this situation is very similar to the electron in the
normalized constant magnetic field. If one rotates the direction of the magnetic
field in any moving frame arbitrarily, 
then he or she will obtain a different set of 
experiments. So, the electron will behave depending on the 
new applied magnetic fields.   
This means that the Bell's inequality associated 
with the physical axes in the rest frame seems to be frame dependent,
which does not obey relativistic covariance.

Finally, we now discuss some subtleties in the calculation of the
Bell's inequality in different works in which Czachor's spin
observable~\cite{czachor} has been used. 
Setting $\vec{a}_1$, $\vec{a}_2$, $\vec{b}_1$, and $\vec{b}_2$ as
Eq.~(\ref{axes:singlet}), the Bell observable
$C(\vec{a}_1,\vec{a}_2,\vec{b}_1,\vec{b}_2)$ for
$|\Psi_{+--+}^{(-)'}\rangle$ is then calculated as
$C'(\vec{a}_1,\vec{a}_2,\vec{b}_1,\vec{b}_2 ; \Psi ')= 2 / \sqrt{2-\beta^2}
(\sqrt{1-\beta^2} + \cos 2 \Omega)$,
which interestingly satisfies the Bell's inequality
$|C(\vec{a}_1,\vec{a}_2,\vec{b}_1,\vec{b}_2)| \to |4 \mathrm{sech}^2
\delta - 2| \le 2$ in the ultrarelativistic limit $\beta \to 1$~\cite{almh},
while it is violated as $|C(\vec{a}_1,\vec{a}_2,\vec{b}_1,\vec{b}_2)|
=2\sqrt{2}$ in the nonrelativistic limit. 
Note that there is a critical value $\beta_c$ satisfying Bell's
inequality, which is somewhat awkward
because it is unnatural in that the Bell's inequality
depends on the boost speed. Fortunately, it has been shown that the maximal
violation of Bell's inequality can be maintained by taking some new
set of spin measurement axes in this calculation~\cite{le}. However,
they are not related to the Lorentz transformation (or the Wigner
rotation) between the two axes.
The origin of this problem is essentially due to the noncovariant definition of
the spin observable given in Ref.~\cite{czachor}.

In conclusion, the Bell observable and the Bell states for Bell's
inequality should be transformed following the principle of
relativistic covariance, which results in a frame-independent Bell's inequality.

\begin{acknowledgments}
We would like to thank C. Ee for helpful discussions. We are also grateful
to M. Czachor for exciting discussions.
This work was supported by Grant No.\ 2000-2-11100-002-5 from the
Basic Research Program of the Korean Science and Engineering
Foundation.
\end{acknowledgments}



\begin{thebibliography}{99}
\bibitem{epr}
A. Einstein, B. Podolsky, and N. Rosen, Phys. Rev. {\bf 47}, 777 (1935)
\bibitem{bell}
J.~S.~Bell, Physics (Long Island City, N.Y.) {\bf 1}, 195 (1964).

\bibitem{czachor}
M.~Czachor, Phys.\ Rev.\ A {\bf 55}, 72 (1997).
\bibitem{am}
P.~M.~Alsing and G.~J.~Milburn, Quantum\ Inf.\ Comput.\ {\bf 2}, 487
(2002).
\bibitem{tu}
H.~Terashima and M.~Ueda, Quantum\ Inf.\ Comput.\ {\bf 3}, 224 (2003);
Int.\ J.\ Quantum\ Inf.\ {\bf 1}, 93 (2003).
\bibitem{almh}
D.~Ahn, H.-j.~Lee, Y.~H.~Moon, and S.~W.~Hwang, Phys.\ Rev.\ A {\bf
  67}, 012103 (2003).
\bibitem{le}
D.~Lee and C.-Y.~Ee, New J.\ Phys.\ {\bf 6}, 67 (2004).
\bibitem{pst}
A.~Peres, P.~F.~Scudo, and D.~R.~Terno, Phys.\ Rev.\ Lett.\ {\bf 88},
230402 (2002); A.~Peres and D.~R.~Terno, Rev.\ Mod.\ Phys.\ {\bf 76},
93 (2004) and references therein.
\bibitem{ohnuki}
Y.~Ohnuki, \textit{Unitary Representations of the Poincare Group and
  Relativistic Wave Equations} (World Scientific, Singapore, 1988).
\bibitem{wigner}
E.~P.~Wigner, Ann.\ Math.\  {\bf 40}, 149 (1939) [Nucl.\ Phys.\ B,
Proc.\ Suppl.\  {\bf 6}, 9 (1989)].
\bibitem{ryder}
L.~H.~Ryder, \textit{Quantum Field Theory}, (Cambridge University
Press, 1986).
\bibitem{nc}
M.~A.~Nielsen and I.~L.~Chuang, \textit{Quantum Computation and
  Quantum information}, (Cambridge University Press, 2000).
\bibitem{chsh}
J.~F.~Clauser, M.~A.~Horne, A.~Shimony, and R.~A.~Holt,
Phys.\ Rev.\ Lett.\  {\bf 23}, 880 (1969).
\end{thebibliography}
\end{document}